\documentclass[journal,10pt]{IEEEtran}
\usepackage{graphicx}
\usepackage{amssymb}
\usepackage{amsmath}
\usepackage{stmaryrd} % Double bracket
\usepackage{amsthm}   % Theorems
\usepackage{multirow} % Table multirow

\theoremstyle{definition}
\newtheorem{lemma}{Lemma}

\newcommand\underrel[2]{\mathrel{\mathop{#1}\limits_{#2}}}
\newcommand\underrelbold[2]{\mbox{\boldmath$\mathrel{\mathop{#1}\limits_{#2}}$}}

\newcommand{\argmin}[1]{\mbox{Arg }\min_{#1}}
\newcommand{\diag}{\mbox{Diag}}

\newcommand{\vE}{\mbox{\boldmath$E$}}

\newcommand{\vQ}{\mbox{\boldmath$Q$}}

\newcommand{\vX}{\mbox{\boldmath$X$}}

\newcommand{\vZ}{\mbox{\boldmath$Z$}}

\newcommand{\vzeros}{{\bf 0}}

\newcommand{\vsigma}{\mbox{\boldmath$\sigma$}}

\newcommand{\vtheta}{\mbox{\boldmath$\theta$}}
\newcommand{\vxi}{\mbox{\boldmath$\xi$}}

\newcommand{\vSigma}{\mbox{\boldmath$\Sigma$}}

\newcommand{\R}{\mathbb{R}}

\newcommand{\T}{\mathbb{T}}

\newcommand{\Z}{\mathbb{Z}}
\newcommand{\prob}{\mbox{Pr}}

\newcommand{\E}{\mbox{E}}

\newcommand{\var}{\mbox{Var}}

\newcommand{\trace}{\mbox{Tr}}

\newcommand{\rA}{\mathcal{A}}

\newcommand{\rD}{\mathcal{D}}

\newcommand{\rF}{\mathcal{F}}
\newcommand{\rG}{\mathcal{G}}

\newcommand{\rI}{\mathcal{I}}
\newcommand{\rM}{\mathcal{M}}

\newcommand{\rR}{\mathcal{R}}
\newcommand{\rN}{\mathcal{N}}

\newcommand{\intEnt}[1]{\llbracket#1\rrbracket}
\newcommand{\ie}{\emph{i.e.}, }
\newcommand{\eg}{\emph{e.g.}, }
\newcommand{\failure}{\rF}
\newcommand{\VAR}[2]{$\mbox{VAR}_{#1}(#2)$}

\begin{document}

\title{Vector autoregressive estimators using binary measurements}

\author{ Colin Cros, Pierre-Olivier Amblard and Jonathan H. Manton, \IEEEmembership{Fellow, IEEE}
\thanks{Copyright (c) 2015 IEEE. Personal use of this material is permitted. However, permission to use this material for any other purposes must be obtained from the IEEE by sending a request to pubs-permissions@ieee.org.}
\thanks{CL and POA are  with  GIPSAlab, Dept DIS, CNRS/Universit\'e Grenoble Alpes/Grenoble INP, France. JM is with the University of Melbourne, Australia.}
\thanks{Manuscript received ; revised }   }% <-this % stops a space

\markboth{IEEE Trans. on Signal Processing,~Vol.~, No.~,~August~2020}{Cros \MakeLowercase{\textit{et al.}}:}  

\maketitle
\begin{abstract}
In this paper, two novel algorithms to estimate a Gaussian Vector Autoregressive (VAR) model from 1-bit measurements are introduced. They are based on the Yule-Walker scheme modified to account for quantisation. The scalar case has been studied before. The main difficulty when going from the scalar to the vector case is how to estimate the ratios of the variances of pairwise components of the VAR model. The first method overcomes this difficulty by requiring the quantisation to be non-symmetric: each component of the VAR model output is replaced by a binary ``zero'' or a binary ``one'' depending on whether its value is greater than a strictly positive threshold.  Different components of the VAR model can have different thresholds. As the choice of these thresholds has a strong influence on the performance, this first method is best suited for applications where the variance of each time series is approximately known prior to choosing the corresponding threshold. The second method relies instead on symmetric quantisations of not only each component of the VAR model but also on the pairwise differences of the components. These additional measurements are equivalent to a ranking of the instantaneous VAR model output, from the smallest component to the largest component. This avoids the need for choosing thresholds but requires additional hardware for quantising the components in pairs. Numerical simulations show the efficiency of both schemes.

\end{abstract}

\IEEEpeerreviewmaketitle

\section{Motivation and overview}

\IEEEPARstart{O}{ne} of the first wireless sensor networks was the ALERT (Automated Local Evaluation in Real Time) system developed in the 1970's by the National Weather Service to record and transmit environmental data~\cite{Arampatzis05}. Since then, such networks have been deployed for different purposes: military \cite{Meesookho02}, environment monitoring \cite{Biagioni02}, agricultural support \cite{Burrel04} and so forth. The proliferation of these systems creates a significant increase in data generation, giving rise to questions of how best to transmit and store data efficiently. Sensors generally have limited available energy (most of the time stored in batteries) and spend most of it sending their data to a command station. Furthermore, the storage of this huge amount of data should be a concern: by 2030, the part of global electricity dedicated to communication technology is expected to jump to 23\% (and up to 50\% in the worst case scenarios, compared to 10\% in 2010) \cite{Anders15}, mainly due to the growing demand for data centers and networks.
Using sensors that produce 1-bit measurements can help reduce the energy burden of transmitting and storing data. This raises the question of how to process such data, and motivates the following specific problem considered in this paper.

%\subsection{Constraints}

Consider a $d$-sensor network \ie a spatially distributed sensor array in which each sensor takes a measurement at regular intervals, \eg a temperature measurement. The sensors transmit their data to a computing station and are able to communicate with other sensors in their vicinity.

Let $\left(Z_i(t)\right)_{t\in\Z}$ be the measurement made by sensor $i$.
We suppose that the measurements  stored in vector $\vZ(t) = \left(Z_1(t), \dots, Z_d(t)\right)^\intercal$ are linked by a Gaussian Vector AutoRegressive (VAR) model.
A $p$-th order Gaussian VAR model, denoted as \VAR{d}{p}, is parameterized by a mean vector $M$, $p$ parameter matrices $\left(A_1, \dots, A_p\right)\in\rM_{d}(\R)^p$ and a covariance matrix $\Sigma_E$. The model is given by
\begin{equation}
    \vZ(t) = M + \sum_{s=1}^p A_s \vZ(t-s) + \vE_t
\end{equation}
where $\left(\vE_t\right)_{t\in\T}$ is a sequence of independent Gaussian vectors with zero mean and covariance $\Sigma_E$.
In the sequel, we will assume that the mean vector $M$ is zero. The unknowns are the $p$ matrices $A_i$ and the covariance $\Sigma_E$.

The aim of the paper is to estimate the parameters of the VAR model from 1-bit quantised data. We will show that the raw data can be compressed into binary measurements without an important loss of precision. \cite{Kedem80} proposed a quite complete study of the scalar autoregressive (SAR) models using binary measurements. The works developed in \cite{Kedem80} on the link between the correlations of a Gaussian process and its binary version paved the way for further results. For example \cite{Krishnamurthy95} showed that adding some noise may be profitable to estimate a scalar AR from binary measurements. The main difficulty that arises when dealing with a vector model is the need to describe the ratios between the time series variance. As a direct consequence the knowledge of the correlations is insufficient to identify the parameters of the process. In contrast, the parameters can be estimated in the scalar case from the sole knowledge of the correlations, the only unobservable parameter being the noise covariance.  For example, consider  two simple \VAR{2}{1} models defined by the set of the following matrices $\left(A_1^1, \Sigma_E^1\right)$ and $\left(A_1^2, \Sigma_E^2\right)$: 
\begin{align*}
    A_1^1 &= \begin{bmatrix}1/2 & 1 \\ 0 & 1/2\end{bmatrix}& \Sigma_E^1 &= \begin{bmatrix}1 & 0 \\ 0 & 1\end{bmatrix} \\
    A_1^2 &= \begin{bmatrix}1/2 & 1/2 \\ 0 & 1/2\end{bmatrix}& \Sigma_E^2 &= \begin{bmatrix}1 & 0 \\ 0 & 2\end{bmatrix}
\end{align*}
The two models can be shown to have the same correlation matrices. However, in the first model, the first component has a bigger variance than the second, whereas in the second model, it is the opposite.

Two algorithms are proposed in section \ref{sec:Scheme1} and \ref{sec:Scheme2} for the estimation of the VAR parameters. They are both adaptations of the well-known Yule-Walker scheme that links the process covariance matrices to its parameters. They both use binary versions of the time series to estimate their correlations. Their main difference comes from the way of estimate the variances of the time series. The first scheme estimates the variance of each time series using a binary version of the process thresholded at a non-null level. The second was designed to be threshold-free, and uses the signs of each time series added to a second binary measurement, a comparison between time series modulus, to estimate the ratios of variances.
Section \ref{sec:Applications} provides some numerical simulations to demonstrate the efficiency of the methods.

\section{Scheme 1}
\label{sec:Scheme1}

This part introduces a simple algorithm to estimate the parameters of a Gaussian VAR model of finite order thanks to a binary version of the process at non-null thresholds. The scheme presented is deduced from the Yule-Walker equations. It requires   to choose thresholds, and this choice induces  a variability in the performance. This scheme is particularly interesting when having some knowledge on the variance of the signals as it requires only $1$ bit per sensor and time-step. 
%An alternative  will be presented in part \ref{sec:Scheme2} to address this issue.

\subsection{Model}
\label{sec:Model}

Let $\vZ(t) = \left(Z_1(t), \dots, Z_d(t)\right)^\intercal$ be a vector of $d$ jointly wide-sense stationary zero-mean real-valued processes. By jointly wide-sense stationary processes, we mean that all the first and second-order moments are defined and invariant by time translation (including second order cross-moments). We further suppose that $\vZ(t)$ follows a Gaussian autoregressive representation of finite order $p$:
\begin{equation}
 \vZ(t) = \sum_{s=1}^p A_s \vZ(t-s) + \vE(t)
\end{equation}
where $\vE(t)$ is a sequence of i.i.d. Gaussian random vectors with  distribution $\rN(\vzeros,\Sigma_E)$. The covariance matrix $\Sigma_E$ and the set of time-invariant matrices $\rA = \left(A_1, \dots, A_p\right)$ are unknown.

As $\vZ(t)$ is wide-sense stationary,  its covariance and correlation matrices can be defined for any integer $\tau$ as:
\begin{align}
    \Gamma(\tau) &= \begin{pmatrix}
        \gamma_{Z_i,Z_j}(\tau)
    \end{pmatrix}_{1\le i,j \le d}  \\
    R(\tau) &= \begin{pmatrix}
        \rho_{Z_i,Z_j}(\tau)
    \end{pmatrix}_{1\le i,j \le d}
\end{align}
with:
\begin{align*}
    \gamma_{Z_i,Z_j}(\tau) &= \E\left[Z_i(t) Z_j(t-\tau)\right] \\
    \rho_{Z_i,Z_j}(\tau) &= \frac{\gamma_{Z_i,Z_j}(\tau)}{\sigma_i \sigma_j}
\end{align*}
where $\sigma_i$ denotes the standard deviation of the $i$-th component of $\vZ(t)$ \ie $\sigma_i = \sqrt{\gamma_{Z_i,Z_i}(0)}$.
If we introduce the diagonal matrix $D_\Gamma = \diag\left(\sigma_1^2, \dots, \sigma_d^2\right)$, the covariance  and the correlation matrices are linked by  $\Gamma(\tau) = D_\Gamma^{\frac{1}{2}}R(\tau)D_\Gamma^{\frac{1}{2}}$.

The process $\vZ(t)$ is completely described by the set of parameters $\vtheta = \left(A_1, \dots, A_p, \Sigma_E\right)$. The aim of the method introduced is to identify this set thanks to a binary version of the process.

\subsection{Yule-Walker Scheme Adaptation}

It is well known that  Yule-Walker equations provide a solution to  the Minimum Least Squares Estimator (MLSE), see \eg \cite{Stine89}. We will show how they are adapted for binary measurements.

The Yule-Walker equations give a recurrence relation between the covariances:
\begin{equation}
    \Gamma(\tau) = \sum_{s=1}^p A_s\Gamma(\tau-s) + \delta_{\tau} \Sigma_E
\end{equation}
where $\delta$ is the Kronecker symbol, $\delta_\tau =1 $ if $\tau=0$ and 0 otherwise.
Considering these equations for $\tau \in \intEnt{0, p}=\{0,\ldots, p\}$ we obtained an invertible system leading to:
\begin{equation}\label{eq:YWeq1}
    A =\Gamma \rG^{-1}
\end{equation}
where $A$, $\Gamma$ are $(d, d(p+1))$-shaped matrices and $\rG$ is $d(p+1)$-shaped square matrix defined as:
\begin{align*}
    A &= \begin{pmatrix}A_1 & A_2 & \cdots & A_p & \Sigma_E \end{pmatrix}\\
    \Gamma &= \begin{pmatrix}\Gamma(1) & \Gamma(2) & \cdots & \Gamma(p) & \Gamma(0)\end{pmatrix}\\
\end{align*}
\begin{equation*}
    \rG = \begin{bmatrix}
        \Gamma(0) & \Gamma(1) & \cdots & \Gamma(p-1) & 0 \\
        \Gamma(1)^\intercal & \Gamma(0) & \cdots & \Gamma(p-2) & 0 \\
        \vdots & \vdots & \ddots & \vdots & \vdots \\
        \Gamma(p-1)^\intercal & \Gamma(p-2)^\intercal & \cdots & \Gamma(0) & 0 \\
        \Gamma(1) &  \Gamma(2) & \cdots & \Gamma(p) & I
    \end{bmatrix}
\end{equation*}

It is more convenient to work with correlations instead of covariances as they can be linked to the binary second order cross-moments {\it via} the generalized arcsine law (cf. lemma \ref{lem:generalizedArcsineLaw}). In the scalar case, multiplying and dividing \eqref{eq:YWeq1} by the variance of the  time series transforms the  covariances of the right-hand side into correlations. In the vector case, the transformation is not that direct as the matrix product is not commutative. 
Let $\rD_\Gamma$ be the $d(p+1)$-shaped diagonal matrix obtained by stacking $p+1$ times matrix $D_\Gamma$:
\begin{equation*}
    \rD_\Gamma = \begin{bmatrix}
    D_\Gamma & 0 & \cdots & 0 \\
    0 & D_\Gamma &  \ddots & \vdots \\
    \vdots &  \ddots & \ddots & 0 \\
    0 & \cdots & 0 & D_\Gamma
    \end{bmatrix}
\end{equation*}
We introduce the correlation matrices $R$ and $\rR$,  derived from the covariance matrices $\Gamma$ and $\rG$, respectively defined by $R = D_\Gamma^{-\frac{1}{2}} \Gamma \rD_\Gamma^{-\frac{1}{2}}$ and $\rR=\rD_\Gamma^{-\frac{1}{2}}\rG\rD_\Gamma^{-\frac{1}{2}}$. Inserting these definitions into \eqref{eq:YWeq1} leads to a relation between the correlations and the parameters of the model and the variances of the process:
\begin{equation}\label{eq:YWeq2}
    \tilde A = \begin{pmatrix}\tilde A_1 & \cdots & \tilde A_p & \tilde \Sigma_E \end{pmatrix} = R \rR^{-1}
\end{equation}
where $\tilde A_s= D_\Gamma^{-\frac{1}{2}} A_s  D_\Gamma^{\frac{1}{2}}$ (\ie the matrices $A_s$ and $\tilde A_s$ are similar) and $\tilde \Sigma_E = D_\Gamma^{-\frac{1}{2}} \Sigma_E D_\Gamma^{-\frac{1}{2}}$.

Formula \eqref{eq:YWeq2} underlines the difficulty of the vector case.  In the scalar case, all parameters except the noise variance are identifiable thanks to the correlations. When dealing with multivariate processes, the correlation matrices only allow to estimate the set of unscaled matrices $\tilde \rA$. In other words, only the diagonal coefficients are identifiable from the correlations. To rescale the parameters, the variance of the processes must be known. Note that in fact,  only the  ratio of the variance are necessary to identify the set $\rA$. If so, the covariance of the noise would be known up to a constant multiplier. We define  $r_{i,j}= \sigma_i / \sigma_j$, the ratio between the $i$-th and the $j$-th time series standard deviations. By denoting as $a_{i,j}^{(s)}$ and $\tilde a_{i,j}^{(s)}$ the entry at the $i$-th row and the $j$-th column in $A_s$ and $\tilde A_s$, we obtain:
\begin{equation*}
    a_{i,j}^{(s)} = r_{i,j} \tilde a_{i,j}^{(s)}
\end{equation*}
If the $r_{i,j}$ are known, $\Sigma_E$ is also identifiable up to a constant multiplier. For example, by denoting as $s_{i,j}$ and $\tilde{s}_{i,j}$ the entry at the $i$-th row and the $j$-th column in $\Sigma_E$ and $\tilde \Sigma_E$, we have:
\begin{equation}
    r_{i,1} r_{j,1} \tilde s_{i,j} = \frac{1}{\sigma_1^2}s_{i,j}.
\end{equation}
Thus, applying this transformation to every entry of $\tilde \Sigma_E$, we obtain $\frac{1}{\sigma_1^2}\Sigma_E$.

\subsection{Estimation scheme}

Let $C = \begin{pmatrix}c_1 & \cdots & c_d\end{pmatrix}^\intercal\in \left(\R^*\right)^d$ be a vector of non-null thresholds. For each component $Z_i(t)$ of the process, a new binary sequence $X_i(t)$ is created by thresholding the process at $c_i$:
\begin{equation}
    X_i(t) = \left[Z_i(t) \ge c_i \right]_I
\end{equation}
where $[\cdot]_I$ denotes the Iverson bracket: $[P]_I = 1$ if $P$ is true,  $[P]_I = 0$ otherwise. We stack these time series into a vector $\vX(t) = \begin{pmatrix}X_1(t) & \cdots & X_d(t)\end{pmatrix}^\intercal$. Obviously, the choice of parameters $c_i$ is significant and will be discussed in part \ref{sec:influence_threshold}. 

To solve the Yule-Walker system, rather than retrieve directly covariances from this binary process, we estimate both the correlations and the variances. The variances are deduced from the empirical mean of the $\vX(t)$'s components while correlations are computed from its covariances. First, for each time series $Z_i(t)$, we associate a standard  threshold $\eta_i=c_i/\sigma_i$ --the term ``standard" is used since $\eta_i$ would have been the threshold if $Z_i(t)$ would have had a standard distribution.

The estimation of the standard thresholds, and consequently of the variances, is straightforward. Since $\Pr(Z_i(t) \ge c_i) = \Phi(-\eta_i)$ we define
\begin{align}
    \hat \eta_i &= -\Phi^{-1}(\bar X_i) \\
    \hat \sigma_i &= \frac{c_i}{\hat \eta_i}
\end{align}
as estimators of the standard threshold and the standard deviation. $\Phi$ denotes the Cumulative Distribution Function (CDF) of a standard Gaussian distribution and $\bar X_i$ the empirical mean of $X_i(t)$.

The estimation of the correlation comes from an adaptation of Sheppard's formula \cite{Sheppard00}:
\begin{multline}
    \prob(Z_1 \ge \eta_1, Z_2 \ge \eta_2) =\\ \frac{1}{2\pi}\int_{\cos^{-1} \rho}^\pi\exp\left(-\frac{\eta_1^2-2\eta_1\eta_2\cos(x)+\eta_2^2}{2\sin^2(x)}\right)dx
\end{multline}
where $(Z_1,Z_2)$ is a bivariate standard Gaussian random variable with correlation $\rho$. This formula links the correlation between the two Gaussian random variables to their binary version's second order cross-moment, a more convenient result is properly stated in lemma \ref{lem:generalizedArcsineLaw}. It is a generalization of the well-known arcsine law and appeared several times in the literature with different demonstrations, see \eg \cite{Owen56}.

\begin{lemma}\label{lem:generalizedArcsineLaw} Let $Z_1$ and $Z_2$ be two standard Gaussian random variables with a correlation $\rho$ and $(\eta_1,\eta_2)\in \R^2$ a couple of thresholds. Then 
\begin{multline}
    \prob(Z_1 \ge \eta_1, Z_2 \ge \eta_2) = \prob(Z_1 \ge \eta_1)\prob(Z_2 \ge \eta_2) +\\
    \int_0^\rho \phi_2(\eta_1,\eta_2 \mid x) dx
\end{multline}
where $\phi_2(\cdot,\cdot\mid \rho)$ denotes the jointly Probability Density Function (PDF) of a bivariate standard Gaussian distribution with a correlation $\rho$.
\end{lemma}
For any couple $(\eta_1,\eta_2)\in \R^2$, we define $\Psi_{\eta_1, \eta_2}: \rho \mapsto \int_0^\rho \phi_2(\eta_1,\eta_2 \mid x) dx$. As $\phi_2$ is a Gaussian PDF, $\Psi_{\eta_1, \eta_2}$ is strictly increasing on $(-1,1)$. Consequently, $\Psi_{\eta_1, \eta_2}$ is invertible for any couple of thresholds, consequently the correlations between the $Z_i$s can be deduced from the binary series covariances:
\begin{equation}\label{eq:rel_binary_variance_continuous_correlation}
    \gamma_{X_i,X_j}(\tau) = \Psi_{\eta_i,\eta_j}\left(\rho_{Z_i,Z_j}(\tau)\right)
\end{equation}
By noting that $\prob(Z_1 \le \eta_1, Z_2 \le \eta_2) = \prob(Z_1 \ge -\eta_1, Z_2 \ge -\eta_2)$, we can extend \eqref{eq:rel_binary_variance_continuous_correlation} to the $0$s events:
\begin{equation}
    \gamma_{1-X_i,1-X_j}(\tau) = \Psi_{\eta_i,\eta_j}\left(\rho_{Z_i,Z_j}(\tau)\right)
\end{equation}
and use:
\begin{equation}\label{eq:rel_binary_variance_continuous_correlation2}
    \rho_{Z_i,Z_j}(\tau) = \Psi_{\eta_i,\eta_j}^{-1}\left(\frac{\gamma_{X_i,X_j}(\tau)+\gamma_{1-X_i,1-X_j}(\tau)}{2}\right)
\end{equation}
Using \eqref{eq:rel_binary_variance_continuous_correlation2} instead of directly inverting \eqref{eq:rel_binary_variance_continuous_correlation} is more reliable as both $0$s and $1$s events are used.

Finally, both variances and correlations can be deduced from the measure $\vX(t)$, and used to estimate the parameters with the following scheme:

\begin{enumerate}
    \item Estimate the standard thresholds \ie compute for $i \in \intEnt{1,d}$:
    \begin{equation*}
        \hat \eta_i = -\Phi^{-1}\left(\bar X_i(t)\right)
    \end{equation*}
    where $\bar X_i(t)$ denotes the empirical mean of $X_i(t)$.
    \item Estimate the covariances of the binary data \ie compute for $i,j \in \intEnt{1,d}$ and $\tau \in \intEnt{0,p}$:
    \begin{multline*}
        \hat \gamma_{X_i,X_j}(\tau)+\hat \gamma_{1-X_i,1-X_j}(\tau) =\\
        \frac{1}{T-\tau}\sum_{t=\tau+1}^T \left[X_i(t) = X_j(t-\tau)\right]_I +\\ -\bar X_i \bar X_j - (1 - \bar X_i)(1 - \bar X_j)
    \end{multline*}
    \item Estimate the correlations between the components \ie compute for $i,j \in \intEnt{1,d}$ and $\tau \in \intEnt{0,p}$:
    \begin{equation*}
        \hat \rho_{Z_i,Z_j}(\tau) = \Psi_{\hat \eta_i, \hat \eta_j}^{-1}\left(\frac{\hat \gamma_{X_i,X_j}(\tau)+\hat \gamma_{1-X_i,1-X_j}(\tau)}{2}\right)
    \end{equation*}
    \item Estimate the parameters $\vtheta$ of the model using the Yule Walker scheme: solve \eqref{eq:YWeq2} to get the estimation of $\tilde \rA$ then rescale the coefficients with the formula:
    \begin{align*}
        \hat  a_{i,j}^{(s)} &= \frac{c_i}{c_j}\frac{\hat \eta_j}{\hat \eta_i} \times \hat{\tilde{a}}_{i,j}^{(s)} \\
        \hat s_{i,j} &= \frac{c_i c_j}{\hat \eta_i \hat \eta_j} \hat{\tilde{s}}_{i,j}
    \end{align*}
     where $a_{i,j}^{(s)}$ (respectively $\tilde a_{i,j}^{(s)}$) denotes the entry at the $i$-th row and  $j$-th column in $A_s$  (resp. $\tilde{A}_s$). Likewise,  $s_{i,j}$ and $\tilde s_{i,j}$ are the entries of $\Sigma_E$ and $\tilde \Sigma_E$.
\end{enumerate}

The main difficulty in the implementation of this algorithm is to compute $\Psi_{\eta_i,\eta_j}^{-1}$. A convenient solution is to first implement $\Psi_{\eta_i,\eta_j}$ using a Gauss-Legendre quadrature and then to compute its inverse using a bisection algorithm as it is strictly increasing. In our implementation, we used three quadratures (from $0$, $1$ and $-1$) with 5 points that we combined to avoid singularity problems in $\pm 1$, as recommended in \cite{Drezner90}.

\subsection{Influence of the thresholds} \label{sec:influence_threshold}

This section discusses the influence of the thresholds on the performance for a number of samples $T$ fixed. It can be easily understood that choosing $c_i$ large (in absolute value) compared to the standard deviation of the process $\sigma_i$ makes all the binary measurements equal ($0$ if $c_i > 0$ or $1$ otherwise), on the other hand choosing $c_i$ small would erase the information about the variances. Therefore, a trade-off must be made as both situations deeply degrade the performance. In the first case, the computation of the empirical means of the $X_i(t)$s will be equal to $0$ or $1$, thus the estimation of the $\hat\eta_i$s would be $\pm \infty$. In the second case, the estimation of the standard deviations becomes both imprecise and inaccurate, leading to a significant loss of precision for the estimation of the non-diagonal parameters of the matrices $A_s$.

Let us first consider the case when $c_i$ is large compared to $\sigma_i$ \ie $\eta_i$ is large (compared to $1$), let us suppose w.l.o.g. that $\forall i \in \intEnt{1,d}$, $c_i > 0$. We denote by ``$\failure$" (for failure) the event ``\emph{the model can not be estimated from the realizations of $\vX(t)$}". We will give a lower bound for $\prob(\failure)$. Several configurations induce ``$\failure$". For instance if the empirical mean of a $X_i(t)$ is $1/2$, then $\hat \eta_i = 0$ and step $4$ can not be performed. In particular, for a given process $i_0$, if the threshold $\eta_{i_0}$ is greater than the maximum value observed among the $T$ realizations, \ie $\forall t\in \intEnt{1,T}$, $X_{i_0}(t)=0$, then the associated standard threshold estimate is $\hat \eta_{i_0} = +\infty$ and the model can not be estimated. We denote as $\rF_{i_0}$ this event: $\rF_i = \bigcap\limits_{t = 1}^T (X_i(t) = 0)$. As $\forall i$, $\rF_i \subset \failure$:
\begin{equation}
    \prob(\failure) \ge \max_{i} \prob(\rF_i)
\end{equation}
$\Pr(\rF_i)$ can be easily bounded:
\begin{align*}
    \prob(\rF_i) &= \prob\left(\bigcap_{t = 1}^T X_i(t) = 0\right) \\
    &= 1 - \prob\left(\bigcup_{t=1}^T X_i(t) = 1\right)
\end{align*}
Yet, the probability of the union is lower than the sum of the probability:
\begin{align*}
    \prob(\rF_i) &\ge 1- T\Phi(-\eta_i) \\
    &\ge 1- \frac{T}{\eta_i \sqrt{2 \pi}} \exp\left(-\frac{\eta_i^2}{2}\right)
\end{align*}
So finally, we can bound the failure probability of the algorithm:
\begin{equation}
    \Pr(\failure) \ge 1 -  \frac{T}{\eta_{\max} \sqrt{2 \pi}} \exp\left(-\frac{\eta_{\max}^2}{2}\right)
\end{equation}
where $\eta_{\max}$ is the maximum of the $\eta_i$s.
The exponential term in the right hand side, requires the thresholds to be at most as large as the standard deviations of the associated time series. For example, for $T=10^4$ samples, if the biggest standard threshold is $\eta_{\max} = 1$, the previous formula does not provide any information: $\prob(\failure) \ge -2400$; but with $\eta_{\max} = 5$ the algorithm will fail with a probability of at least $99.7\%$.

Now, we consider the case where $c_i$ is small compared to $\sigma_i$ \ie $\eta_i$ is small (compared to 1). The estimation of $r_{i,j}$ is obtained from the estimation of both $\sigma_i$ and $\sigma_j$ which are estimated using:
\begin{equation*}
    \hat \sigma_i = \frac{c_i}{\hat \eta_i} = \frac{-c_i}{\Phi^{-1}\left(\bar X_i\right)}
\end{equation*}
The estimation of the ratio can be expressed as a function of the true ratio, the standard thresholds and their estimates:
\begin{align*}
    \hat r_{i,j} &= \frac{\hat \sigma_i}{\hat \sigma_j} = \frac{c_i}{c_j} \frac{\hat \eta_j}{\hat \eta_i} = r_{i,j} \frac{\eta_i}{\eta_j} \frac{\hat \eta_j}{\hat \eta_i} = r_{i,j} \frac{\eta_i}{\eta_j} \frac{\Phi^{-1}(\bar X_j)}{\Phi^{-1}(\bar X_i)}
\end{align*}
Note that $\bar X_i$ depends on the threshold $c_i$, and for the rest of this section, we will note $\bar X_i(c_i)$ to highlight this. To ensure that the ratio is well defined, we used an odd number of samples, thus $|\hat \eta_i|\ge\Phi^{-1}\left(\frac{1}{2}+\frac{1}{2T}\right)$. To understand the behaviour when the thresholds are small, we must focus on the behaviour of the ratio
\begin{equation*}
    \frac{\eta_i}{\eta_j} \frac{\Phi^{-1}(\bar X_j(c_j))}{\Phi^{-1}(\bar X_i(c_i))}
\end{equation*}
If the number of samples is great enough we can neglect the events $(\bar X_k = 0)$ and $(\bar X_k = 1)$ which are very unlikely as they occur only if all the binary measurements are equal. Let us note $\delta_k = \prob((\bar X_k = 1) \cup (\bar X_k = 0))$, we have with a probability of at least $1-\delta_i-\delta_j$:
\begin{equation*}
    \left|\frac{\eta_i}{\eta_j} \right|\frac{\Phi^{-1}\left(\frac{1}{2}+\frac{1}{2T}\right)}{\Phi^{-1}\left(1-\frac{1}{T}\right)} \le 
    \left| \frac{\eta_i}{\eta_j} \frac{\Phi^{-1}(\bar X_j(c_j))}{\Phi^{-1}(\bar X_i(c_i))} \right| \le
    \left|\frac{\eta_i}{\eta_j} \right|\frac{\Phi^{-1}\left(1-\frac{1}{T}\right)}{\Phi^{-1}\left(\frac{1}{2}+\frac{1}{2T}\right)}
\end{equation*}
If $T$ is large we can use the asymptotic equivalents: $\Phi^{-1}\left(1-\frac{1}{T}\right) \sim \sqrt{2\ln T}$ and $\Phi^{-1}\left(\frac{1}{2}+\frac{1}{2T}\right) \sim \frac{\sqrt{2\pi}}{2T}$:
\begin{equation*}
    \left|\frac{\eta_i}{\eta_j} \right| \frac{\sqrt{\pi}}{2T\sqrt{\ln T}} \le
    \left| \frac{\eta_i}{\eta_j} \frac{\Phi^{-1}(\bar X_j(c_j))}{\Phi^{-1}(\bar X_i(c_i))} \right| \le
    \left|\frac{\eta_i}{\eta_j} \right| \frac{2T\sqrt{\ln T}}{\sqrt{\pi}}
\end{equation*}
As a consequence, if $\eta_i$ tends to $0$ with $\eta_j$ fixed (or $\eta_j = o(\eta_i)$), the estimation of the ratio would tends to $0$ with a probability of at least $1-\delta_i-\delta_j$; if on the other hand $\eta_j$ tends to $0$ with $\eta_i$ fixed (or $\eta_i = o(\eta_j)$) the estimation of the ratio can explode.
The case where both $\eta_i$ and $\eta_j$ tend to 0 with a ratio $\eta_i/\eta_j$ constant is more difficult. Even if we exclude the events where the ratios are not defined and state that $\Phi^{-1}(\bar X_k(c_k))$ converges to $\Phi^{-1}(\bar X_k(0))$, the ratio estimator converges to a Cauchy distribution when $T \to \infty$.

\subsection{Discussion}

The algorithm introduced in this part, allows to completely identify a Gaussian VAR using a binary version of it. From a computational point of view, except for the implementation of $\Psi_{\eta_1,\eta_2}^{-1}$ that may need some further research, the method is similar to the usual Yule-Walker scheme. The use of binary sequences simplify the implementation of the moments. The sensors only need to send $1$-bit measurements which makes this algorithm very data efficient. This algorithm fits with applications where we have some \emph{a priori} expectation for the variance of the process, it would be interesting to develop adaptive schemes to optimize the thresholds. If the sensors are able to run some basic calculations, another solution could be to make the sensors send only the sign of the process (\ie set the thresholds to $0$) then, at the end of the record send the empirical variance computed from the raw measurements. If the number of samples is great enough, sending a continuous value would be negligible.

When the thresholds are set to $0$, the variances cannot be estimated from the measurements. However, if only the set of parameters $\rA$ is to be estimated, only the ratios between variances have to be identify, so another kind of threshold-free binary measure can be designed to retrieve them. The following part introduces such a method.

\section{Scheme 2}
\label{sec:Scheme2}

This part presents a threshold-free method to estimate the set of matrices $\rA$. This scheme relies on the measure of the sign of the process and a new measure to retrieve the ratios between the variances of the process. First, we discuss the model identifiability solely from the ratios and the correlations, then we present the measures used to estimate them and how they fit into the Yule-Walker scheme.

\subsection{Model identifiability}

The aim of the algorithm introduced here is to identify the set of matrices $\rA$. In part \ref{sec:Model}, we explained that in addition to the correlations, we need to know the ratios between the variances of the components. If we can estimate these two aspects of the process, we can also deduce the shape of the covariance matrix of the noise (we saw that we can find a matrix proportional to it).

However the noise power cannot be identified as it is intrinsically linked to the variance of the process. Multiplying by a constant the noise covariance  only rescales the realizations of the process, but does not alter the relations between the time series. This is straightforward when considering the moving average representation of the process $\vZ(t) = \sum_{s=0}^{+\infty}B_s \vE(t-s)$. For this reason, we choose to group VAR models in equivalence classes defined by the set of matrices $\rA$ and a covariance matrix $\Sigma_E^0$ satisfying $\trace(\Sigma_E^0) = d$. We denote  as $\tilde \vtheta = (A_1, \dots, A_p, \Sigma_E^0)$  such a class. A VAR model $\vtheta = (A_1', \dots, A_p', \Sigma_E)$ belongs to the class $\tilde \vtheta$ if the sets of matrices $\rA$ and $\rA'$ are equal and if $\Sigma_E^0$ is proportional to $\Sigma_E$. From the ratios and the correlations only the class of the model is identifiable.

\subsection{Measures}

As in the preceding scheme, for each time series $Z_i(t)$, a binary time series $X_{i}(t)$ is defined by thresholding $Z_i(t)$ but at $0$:
\begin{equation}
    X_{i}(t) = \left[Z_i(t) \ge 0\right]_I
\end{equation}
These time series are once again stacked into vector $\vX(t)$.

Moreover, in order to identify the ratios between two components' variance, we have chosen to record for every couple $(Z_i,Z_j)$ the maximum in absolute value:
\begin{equation}
    Q_{i,j}(t) = \left[ |Z_i(t)| \ge |Z_j(t)| \right]_I
\end{equation}
The larger $\sigma_i$ compared to $\sigma_j$, the more likely $\left[Q_{i,j}(t) = 1\right]$ --in other words, $Q_{i,j}(t)$ records the dominance of the $i$-th time series over the $j$-th time series. We stack  all these couples in a  $d(d-1)/2$-shaped vector $\vQ(t)$.
% 
% POA/ but this requires at first glance that avery sensor communicates with everyone. Comments.
With these two measures, at each time step, $d(d+1)/2$ measurements are recorded. We will see however that only $2d-1$ are really necessary.

\subsection{Estimation scheme}

To use the Yule Walker equations, the correlations and the variance ratios must be retrieved from the binary measurements. The correlations are deduced from the transition probabilities $\lambda_{i,j}(\tau)$ defined as:
\begin{equation}
    \lambda_{i,j}(\tau) = \prob\left(X_i(t) = 1 \mid X_j(t-\tau) = 1\right)
\end{equation}
These probabilities are estimated using the binary measure $\vX(t)$ and the extension to vectors of the maximum likelihood estimator introduced by Kedem in \cite{Kedem80}. The ratios are deduced from the second measure $\vQ(t)$. To link both measures to the relevant information, the two following lemmas are used. The first is the very well-known arcsine law, dating back to Sheppard \cite{Sheppard00}, and rediscovered many times later, see {\it e.g.} \cite{VanVM66}. We found no trace of the second in the literature, and its simple proof is given in section \ref{proof:predominace}.

\begin{lemma}\label{lem:arcsineLaw} Let $Z_1$ and $Z_2$ be two centered Gaussian random variables with correlation $\rho$. Then
\begin{equation}
    \prob(Z_1 \ge 0,Z_2 \ge0) = \frac{1}{4} + \frac{1}{2\pi}\arcsin \rho
\end{equation}
\end{lemma}

\begin{lemma}\label{lem:ratios} Let $Z_1$ and $Z_2$ be two centered Gaussian random variables with correlation $\rho$ and variances $\sigma_1^2$ and $\sigma_2^2$. Then
\begin{multline}
   \prob( |Z_1| \ge |Z_2|) = \frac{1}{\pi} \left[\arctan\left(\frac{\frac{\sigma_1}{\sigma_2}-\rho}{\sqrt{1-\rho^2}}\right)\right. \\
   + \left.\arctan\left(\frac{\frac{\sigma_1}{\sigma_2}+\rho}{\sqrt{1-\rho^2}}\right)\right]    
\end{multline}
\end{lemma}

Lemma \ref{lem:arcsineLaw} is just a special case of lemma \ref{lem:generalizedArcsineLaw} with both thresholds equal to $0$; this relation is nevertheless much easier to invert. The correlation can be deduced from $\prob\left[Z_i(t) \ge 0, Z_j(t-\tau) \ge 0\right]$ or equivalently from $\lambda_{i,j}(\tau)$ by:
\begin{equation}
    \rho_{Z_i,Z_j}(\tau) = \sin\left[ \pi\left(\lambda_{i,j}(\tau)-\frac{1}{2}\right)\right]
\end{equation}
 Kedem gave a maximum likelihood estimator for the transition probabilities assuming wrongly (but on purpose) that the binary process $\vX(t)$ is Markovian. Even if $\vZ(t)$ is a $p$-th order Markov process, $\vX(t)$ is not.  Assuming that $T$ consecutive observations of $\vX(t)$ are recorded, and using the Markov hypothesis, Kedem's maximum likelihood estimator becomes in the vector case:
\begin{equation}
    \hat \lambda_{i,j}(\tau) = \frac{2R_{i,j}(\tau) - (S_i+S_j) + T }{T-\tau}
\end{equation}
where $R_{i,j}(\tau) = \sum_{t=\tau+1}^T X_i(t)X_j(t-\tau)$ and for $k \in \{i,j\}$, $S_k = \sum_{t=1}^T X_k(t)$. If we neglect the edge effects, the MLE can be reformulated as: 
\begin{equation}\label{eq:transition_proba_MLE2}
    \hat \lambda_{i,j}(\tau) = \frac{1}{T-\tau} \sum_{t=\tau+1}^T \left[X_i(t) = X_j(t-\tau)\right]_I
\end{equation}
This formula  is easier to understand: the transition probability is approximated by the average number of samples for which $Z_i(t)$ and $Z_j(t-\tau)$ have the same sign. Furthermore, this estimator is straightforward to compute and takes credit from both $0$s and $1$s measurements, even though $\lambda_{i,j}(\tau)$ is only concerned with  $1$s events. Thus all measurements contribute to the estimator defined by \eqref{eq:transition_proba_MLE2}.

Lemma \ref{lem:ratios} spells out the link between the second measure $\vQ(t)$ and the ratios between components' variance. The larger the ratio $r_{i,j}$, the larger the likelihood that the magnitude of $Z_i(t)$ is larger than the magnitude of $Z_j(t)$. As for the other lemmas, the relation can be inverted {\it via}:
\begin{equation} \label{eq:PredomInverse}
   \frac{\sigma_1}{\sigma_2} = \sqrt{\frac{1-\rho^2}{\tan^2 \pi p} + 1} - \frac{\sqrt{1-\rho^2}}{\tan \pi p}
\end{equation}
where $p=\prob(|Z_1| \ge |Z_2|)$. 
Consequently, the ratio $r_{i,j}$ can be retrieved from the correlation and the probability of preponderance.

The estimation scheme follows, and can be summed up as:
\begin{enumerate}
    \item Estimate the transition probabilities using the MLE \ie compute for $i,j \in \intEnt{1,d}$ and $\tau \in \intEnt{0,p}$: \begin{equation*}
        \hat \lambda_{i,j}(\tau) = \frac{1}{T-\tau} \sum_{t=\tau+1}^T \left[X_i(t) = X_j(t-\tau)\right]_I
    \end{equation*}
    \item Estimate the correlations between the time series $Z_i$ using the arcsine law \ie compute for $i,j \in \intEnt{1,d}$ and $\tau \in \intEnt{0,p}$:
    \begin{equation*}
        \hat \rho_{Z_i,Z_j}(\tau) = \sin\left[\pi\left( \hat \lambda_{i,j}(\tau) - \frac{1}{2}\right)\right]
    \end{equation*}
    \item Estimate the ratios $r_{i,j}$ using \eqref{eq:PredomInverse}  \ie compute for $i,j \in \intEnt{1,d}$:
    \begin{align*}
        \bar Q_{i,j} &= \frac{1}{T}\sum_{t=1}^T Q_{i,j}(t) \\
        \hat r_{i,j} &= \sqrt{\frac{1-\hat \rho_{Z_i,Z_j}(0)^2}{\tan^2 \pi \bar Q_{i,j}} + 1} - \frac{\sqrt{1-\hat \rho_{Z_i,Z_j}(0)^2}}{\tan \pi \bar Q_{i,j}}
    \end{align*}
    \item Estimate the class of the model $\tilde \vtheta$ using the Yule Walker scheme: solve \eqref{eq:YWeq2} to get the estimation of $\tilde A$, and then rescale the coefficients obtained:
    \begin{align*}
        \hat  a_{i,j}^{(s)} &= \hat r_{i,j} \times \hat{\tilde{a}}_{i,j}^{(s)} \\
        \hat s_{i,j}  &= \hat r_{i,1} \hat r_{j,1} \hat{\tilde{s}}_{i,j}
    \end{align*}
    where $a_{i,j}^{(s)}$ (respectively $\tilde a_{i,j}^{(s)}$) denotes the entry at the $i$-th row and  $j$-th column in $A_s$  (resp. $\tilde{A}_s$). Likewise,  $s_{i,j}$ and $\tilde s_{i,j}$ are the entries of $\Sigma_E$ and $\tilde \Sigma_E$. Deduce the class of $\vtheta$ by dividing  the noise covariance estimation by its trace.
\end{enumerate}

The measure $\vQ(t)$ records some redundant information. As $r_{i,j}=\sigma_i/\sigma_j$, $d-1$ measures are sufficient to estimate all these ratios. For example, measuring only  $Q_{1,j}(t)$ for $j \in \intEnt{2,d}$ is sufficient to deduce the whole set of ratios, since $r_{i,j}=r_{1,j}/r_{1,i}$. This approach has the advantage of requiring as few data as possible since it presents no redundancy in the measures. On the other hand, it implies a loss of precision as two inaccurate estimations are combined to form another one, and the precision induced is not homogeneous: the estimation would be more accurate on the coefficients of the first row of the matrices. If a such drastic reduction of data is not needed, it is still possible to take advantage of this redundancy. As stated before, the vector of variances $\vSigma_{Z} = (\sigma_1^2,\dots, \sigma_d^2)^\intercal$ can not be deduced from the measures $\vQ(t)$, only a vector proportional to it $\vSigma_{Z}'\propto \vSigma_{Z}$ is identifiable. By imposing the norm of  $\vSigma_{Z}$ to be equal to a constant, the vector $\vSigma_Z'$ can be estimated from the $r_{i,j}$ using an ordinary least-square optimization. Doing so, the whole set of data is used to estimate each ratio. Note that this technique is effective only for $d\ge4$. Indeed, it is effective only if the number of measures is greater than the number of unknown factors. For $d = 3$ it just homogenizes the error of the three measures on the three ratio estimations.

These three approaches are further discussed in part \ref{sec:Applications}.

\subsection{Discussion}

The second scheme presented in this part provides a threshold-free method to identify the model up to a constant multiplier. It can apply without any \emph{a priori} information on the data. In practice, it requires the sensors to be interconnected to compare the measurements but an interesting point is that the sensor network does not have to be fully connected: the graph $\rG = (V,E)$ formed by the sensors only needs to be connected. However, the closer two sensors in the graph, the more accurate the ratio's estimation, so a complete graph would achieve better performances. A balanced must be found between the completeness of the graph and the precision: on the one hand, the more connected the graph, the more data are sent: the number of bits transmitted at each time step is $d+|E|$ where $|E|$ is the number of edges in $\rG$; on the other hand, the variance of the estimation of a parameter $a_{i,j}$ increases with the distance in $\rG$ between the $i$-th and $j$-th sensor. This variance can be decomposed, under reasonable independence assumptions (cf. part \ref{proof:graph_representation}), in the sum of a constant term and a term function of the distance in the graph between the $i$-th and the $j$-th sensor:
\begin{equation}
    \var(\hat a_{i,j}) = r_{i,j}^2\sigma_{\tilde a}^2 + f(i,j) \sigma_r^2
\end{equation}
where $\sigma_{\tilde a}^2$, $\sigma_r^2$ are the variances of the estimation of the unscaled parameters and the ratios (assumed constant for all parameters to simplify notations) and $f$ is the function which depends on the graph and increases with the distance - see part \ref{proof:graph_representation}.

\section{Applications}\label{sec:Applications}

In this section, the performance of the algorithms presented are compared. We denote as \VAR{d}{p} a vector autoregressive model of order $p$ with $d$ components.
First, we will analyze the loss in precision induced by the quantification on \VAR{2}{1} models--the simplest non-scalar VAR--for both methods. Then we will focus on higher dimension \VAR{d}{1} models and especially how the redundancy in $\vQ(t)$ can be exploited to either increase the precision or reduce the amount of data.

We shall not discuss higher order cases as any \VAR{d}{p} can be reduced as a \VAR{pd}{1}:
\begin{equation}
    \underrelbold{Z}{\sim}(t) = \underrel{A}{\sim} \underrelbold{Z}{\sim}(t-1) + \underrelbold{E}{\sim}(t)
\end{equation}
where $\underrelbold{Z}{\sim}(t)$ is obtained by stacking $\vZ(t)$ and its $p-1$ preceding realizations, $\underrel{A}{\sim}$ is the companion block matrix associated to the system and $\underrelbold{E}{\sim}(t)$ is the extended noise vector:
\begin{align*}
    \underrelbold{Z}{\sim}(t) &= \begin{pmatrix} \vZ(t)^\intercal & \vZ(t-1)^\intercal & \cdots & \vZ(t-p+1)^\intercal \end{pmatrix}^\intercal\\
    \underrelbold{E}{\sim}(t) &= \begin{pmatrix} \vE(t)^\intercal & 0 & \cdots & 0 \end{pmatrix}^\intercal \\
    \underrel{A}{\sim} &= \begin{bmatrix}
        A_1 & A_2 & \cdots & A_{p} \\
        I_d & 0 & \cdots & 0 \\
        0 & \ddots & \ddots & \vdots \\
        0 & 0 & I_d & 0
    \end{bmatrix}
\end{align*}

\subsection{The simplest example: \VAR{2}{1}}

\subsubsection{Scheme 1} 
\begin{figure}
\begin{center}
    \includegraphics[scale=.35]{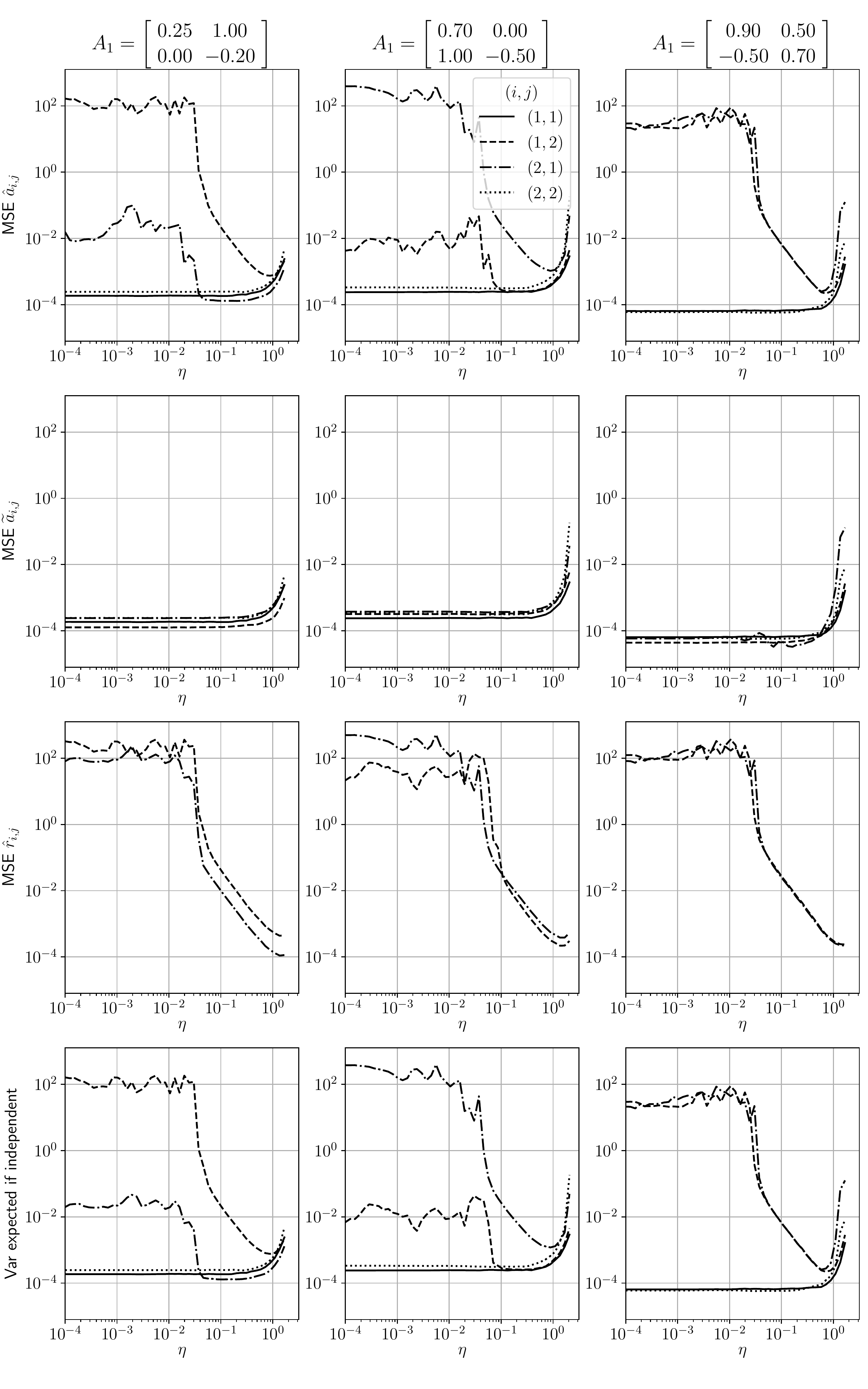}
    \end{center}
        \caption{MSE of the first estimator function of the standard thresholds for three \VAR{2}{1}. For each column, the matrix $A_1$ used is indicated on top of it.  Both normal thresholds $\eta_1$ and $\eta_2$ are variable and equal, their value is given by the X-axis. First row: MSE of the estimated coefficients of $A_1$. Second raw: MSE of the estimated coefficients of the unscaled matrix $\tilde A_1$. Third raw: MSE of the estimated ratios $\sigma_i / \sigma_j$. Fourth raw: Variance expected if the estimation of the unscaled matrix and the ratio would have been independent. For all the simulations, $10^4$ samples were used and the MSE have been average on $1001$ realizations.} 
\label{fig:non_parametric_VAR2_1_both_eta}
\end{figure}

\begin{figure}
\begin{center}
    \includegraphics[scale=.35]{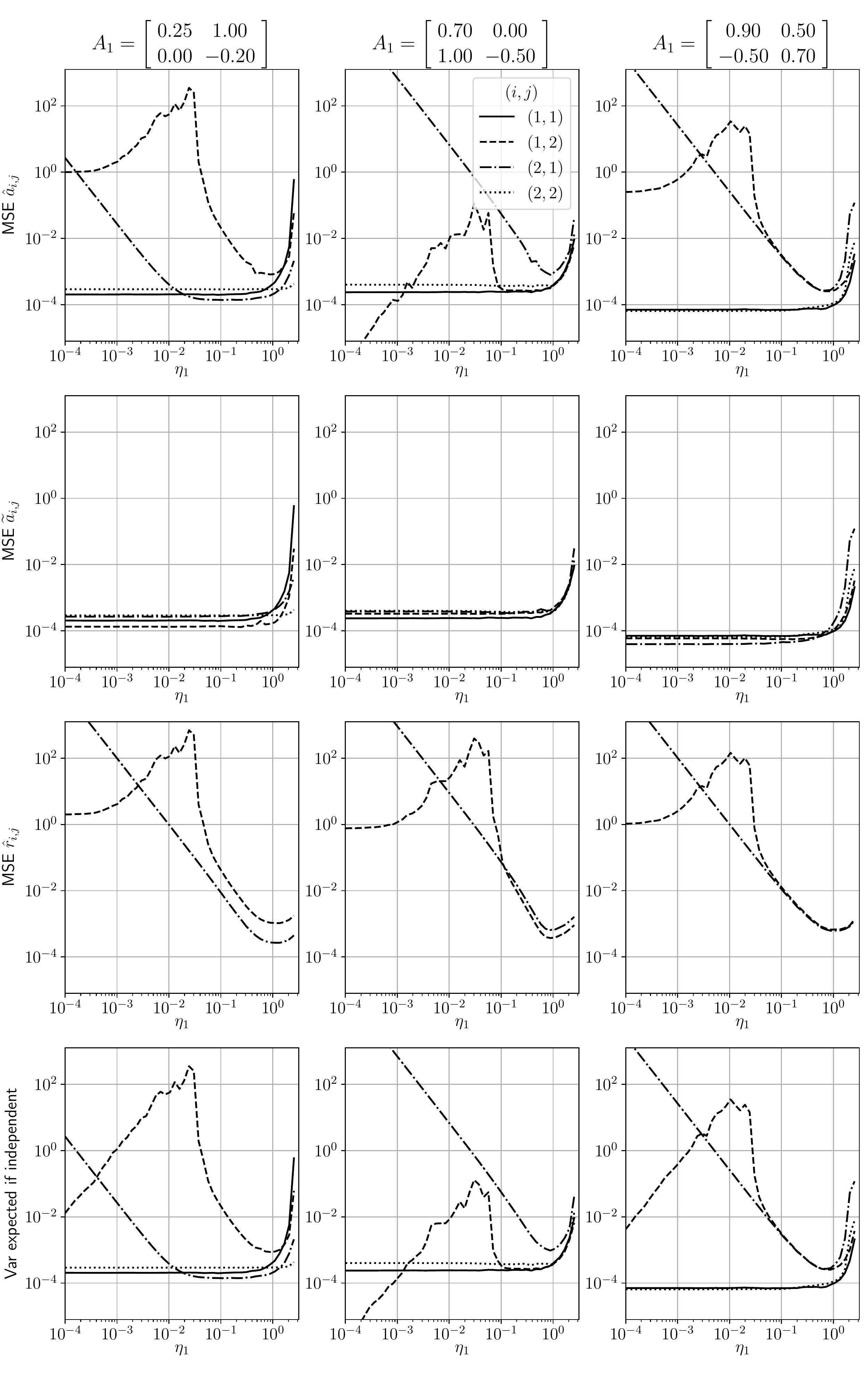}
    \end{center}
        \caption{MSE of the first estimator function of the standard threshold $\eta_1$ for three \VAR{2}{1}. For each column, the matrix $A_1$ used is indicated on top of it. Only $\eta_1$ is variable, its value is given by the X-axis, $\eta_2$ is fixed equal to $0.5$. First raw: MSE of the estimated coefficients of $A_1$. Second raw: MSE of the estimated coefficients of the unscaled matrix $\tilde A_1$. Third raw: MSE of the estimated ratios $\sigma_i / \sigma_j$. Fourth raw: Variance expected if the estimation of the unscaled matrix and the ratio would have been independent. For all the simulations, $10^4$ samples were used and the MSE have been average on $1001$ realizations.} 
\label{fig:non_parametric_VAR2_1_eta1}
\end{figure}

Figures \ref{fig:non_parametric_VAR2_1_both_eta} and \ref{fig:non_parametric_VAR2_1_eta1} present the Mean Square Error (MSE) of the estimator using thresholds for three different \VAR{2}{1} and a large range of thresholds. The parameters are given in the figure. In figure \ref{fig:non_parametric_VAR2_1_both_eta} the two standard thresholds are equal and variable whereas in figure \ref{fig:non_parametric_VAR2_1_eta1} only the first threshold is variable. The performance of this scheme are very heterogeneous between the diagonal and non-diagonal parameters. We can note that for the diagonal coefficients, the MSE becomes constant as the $\eta_i$s tend to 0. This is easily understandable since the estimation of a diagonal coefficient corresponds to the influence of one time series on itself and, consequently, does not depend on any ratio of variances; it can be determined even when all the standard thresholds are null. For the non-diagonal coefficients, when both $\eta_1$ and $\eta_2$ tend to $0$ (first raw in fig. \ref{fig:non_parametric_VAR2_1_both_eta}), the MSE significantly increases with a different magnitude whether the coefficient $a_{i,j}$ is null or not. We can explain this phenomenon by assuming that the estimations of the ratio and the unscaled matrices are independent for small thresholds. This assumption makes sense as correlations and ratio record two intrinsically different aspects of the model: correlations measure the sign concordance whereas the ratio measure the predominance of one series over another. Knowing that $|Z_i(t)|$ is greater than $|Z_j(t)|$ does not provide any information on the sign of $Z_i(t)$ and $Z_j(t)$ and conversely, knowing whether $Z_i(t)$ and $Z_j(t)$ have the same sign does not indicate which one is the greater. Under this assumption, the variance of the estimator of $a_{i,j}^{(s)}$, the entry at the $i$-th row and $j$-th column of $A_s$, is:
\begin{multline}\label{eq:VAR_indep}
    \var\left(\hat a_{i,j}^{(s)}\right) = \var\left(\hat{\tilde{a}}_{i,j}^{(s)}\right)\var\left(\hat r_{i,j}\right) +\\
    \var\left(\hat{\tilde{a}}_{i,j}^{(s)}\right)\E\left(\hat r_{i,j}\right)^2 + \var\left(\hat r_{i,j}\right) \left.\tilde a_{i,j}^{(s)}\right.^2
\end{multline}
This assumption is justified by the fourth row in figure \ref{fig:non_parametric_VAR2_1_both_eta} which presents the expected variance of the estimator under this assumption. Consequently the bigger $|\tilde{a}_{i,j}^{(s)}|$, the bigger the variance of $\hat{a}_{i,j}^{(s)}$.
Another interesting point is that the MSE on the unscaled coefficients (second raw) remains almost constant on a large range of values and explodes from $\eta_i = 1$ leading to failure of the computation for $\eta_i \ge 5$. The MSE of the estimation of the ratio (third raw) becomes very bad as the thresholds tend to $0$ --since the variances can not be estimated with a null thresholding level. The behaviours described in part \ref{sec:influence_threshold} are confirmed: when $\eta_i$ is fixed and $\eta_j$ goes to $0$, the variance of $\hat r_{i,j}$ is proportional to $1/\eta_j^2$ (fig. \ref{fig:non_parametric_VAR2_1_eta1}, $\hat r_{2,1}$); when $\eta_j$ is fixed and $\eta_i$ tends to $0$ (fig. \ref{fig:non_parametric_VAR2_1_eta1}, $\hat r_{1,2}$), we can guess that the estimation of $a_{i,j}$ tends to 0 as the MSE becomes constant only if $a_{i,j} \ne 0$; when both $\eta_i$ and $\eta_j$ (fig. \ref{fig:non_parametric_VAR2_1_both_eta}, both $\hat r_{1,2}$ and $\hat r_{2,1}$) tend to 0, we notice that the MSE increases before tending to a constant which suggests that the estimation is biased.

\subsubsection{Scheme 2}
Table \ref{tab:VAR2-1_perf} compares the performance of the threshold-free estimator (denoted as $\hat A_1$) to the Minimum Least Squares Error Estimator (MLSE) on the continuous data (denoted as $\bar A_1$) for the same \VAR{2}{1} and different numbers of samples. We can observe that in average both estimations have the same accuracy, that tends to justify the use of our estimator. The MLSE has nevertheless a smaller Mean Square Error (MSE) --about $4.4$ time smaller--which confirms the intuition that thresholding a continuous process implies a loss of information. We can wonder if the loss of precision comes more from the estimation of the correlations or the ratio of standard deviations. Table \ref{tab:VAR2-1_perf2} exposes the influences of these different estimations on the error, providing the MSE on the unscaled matrix $\tilde A_1$ and the ratios. We can note that these two MSE are of the same order of magnitude, this means that estimating the ratios of standard deviations does not imply a high loss of precision on the non-diagonal parameters. The last column gives the expected variance for the estimation of $\hat A_1$ if the two preceding estimation would truly have been independent. This expected value, computed using \eqref{eq:VAR_indep}, is very close to the empirical variance, which tends to confirm the assumption. Finally, we can observe that when doubling $T$, the MSE is divided by two, which suggests that the asymptotic variances of the estimators are proportional to $1/T$. This result was well known by Kedem who provided asymptotic distributions for the scalar estimators.

\begin{table*}[t]
\centering
\caption{Average and mean squared error of the threshold-free estimator and MLSE (on the continuous data) for three \VAR{2}{1} and different numbers of samples. For all the configurations, the input noise has a standard Gaussian distribution and $10^3$ simulations were made to compute the average and MSE matrices.}
\label{tab:VAR2-1_perf}
\begin{tabular}{|c|r|r r|r r|r r|r r|}
    \hline\rule{0em}{1.25em}
    $A_1$ & $T$ & \multicolumn{2}{c|}{ave. $\hat A_1$} &\multicolumn{2}{c|}{ave. $\bar A_1$} &\multicolumn{2}{c|}{MSE $\hat A_1$} &\multicolumn{2}{c|}{MSE $\bar A_1$}\\
    \hline
    \multirow{8}{*}{$\begin{pmatrix} 0.25 & 1.00 \\ 0.00 & -0.20 \end{pmatrix}$}& 250 & 0.2476 & 0.9975 & 0.2450 & 1.0006 & 7.9e-03 & 1.7e-02 & 2.1e-03 & 4.1e-03 \\ & & -0.0029 & -0.2041 & -0.0048 & -0.2046 & 5.8e-03 & 1.1e-02 & 2.1e-03 & 4.0e-03\\ \cline{2-10}
     & 500 & 0.2479 & 0.9976 & 0.2474 & 0.9999 & 4.0e-03 & 8.2e-03 & 1.1e-03 & 2.0e-03 \\ & & -0.0023 & -0.2008 & -0.0024 & -0.2016 & 2.8e-03 & 5.2e-03 & 1.1e-03 & 2.0e-03\\ \cline{2-10}
     & 1000 & 0.2482 & 0.9972 & 0.2483 & 0.9992 & 1.9e-03 & 4.3e-03 & 5.3e-04 & 9.9e-04 \\ & & -0.0014 & -0.2009 & -0.0012 & -0.2017 & 1.3e-03 & 2.4e-03 & 5.4e-04 & 9.6e-04\\ \cline{2-10}
     & 2000 & 0.2493 & 0.9975 & 0.2496 & 0.9995 & 9.4e-04 & 2.1e-03 & 2.6e-04 & 4.7e-04 \\ & & -0.0005 & -0.2000 & -0.0004 & -0.2010 & 6.8e-04 & 1.1e-03 & 2.5e-04 & 4.7e-04\\ 
    \hline
    \multirow{8}{*}{$\begin{pmatrix} 0.70 & 0.00 \\ 1.00 & -0.50 \end{pmatrix}$}& 250 & 0.7019 & -0.0051 & 0.6904 & -0.0034 & 8.7e-03 & 9.6e-03 & 2.6e-03 & 2.0e-03 \\ & & 0.9982 & -0.4971 & 0.9984 & -0.4966 & 1.7e-02 & 1.0e-02 & 2.8e-03 & 2.0e-03\\ \cline{2-10}
     & 500 & 0.7041 & -0.0072 & 0.6950 & -0.0017 & 4.3e-03 & 4.7e-03 & 1.3e-03 & 9.8e-04 \\ & & 0.9993 & -0.5009 & 0.9994 & -0.4987 & 7.6e-03 & 5.2e-03 & 1.3e-03 & 1.0e-03\\ \cline{2-10}
     & 1000 & 0.7026 & -0.0049 & 0.6973 & -0.0010 & 2.1e-03 & 2.2e-03 & 6.3e-04 & 4.7e-04 \\ & & 1.0005 & -0.5016 & 0.9992 & -0.4988 & 3.7e-03 & 2.5e-03 & 6.6e-04 & 5.1e-04\\ \cline{2-10}
     & 2000 & 0.7002 & -0.0013 & 0.6982 & -0.0004 & 1.1e-03 & 1.1e-03 & 3.3e-04 & 2.4e-04 \\ & & 1.0003 & -0.5008 & 0.9995 & -0.4989 & 1.9e-03 & 1.2e-03 & 3.3e-04 & 2.5e-04\\ 
    \hline
    \multirow{8}{*}{$\begin{pmatrix} 0.90 & 0.50 \\ -0.50 & 0.70 \end{pmatrix}$}& 250 & 0.8985 & 0.5022 & 0.8980 & 0.5005 & 2.6e-03 & 3.1e-03 & 5.2e-04 & 4.9e-04 \\ & & -0.5014 & 0.7011 & -0.5001 & 0.6970 & 2.6e-03 & 2.2e-03 & 5.1e-04 & 5.2e-04\\ \cline{2-10}
     & 500 & 0.8993 & 0.4997 & 0.8992 & 0.4999 & 1.3e-03 & 1.5e-03 & 2.5e-04 & 2.4e-04 \\ & & -0.5013 & 0.7015 & -0.4999 & 0.6980 & 1.2e-03 & 1.1e-03 & 2.6e-04 & 2.6e-04\\ \cline{2-10}
     & 1000 & 0.8995 & 0.5002 & 0.8995 & 0.5000 & 6.0e-04 & 7.7e-04 & 1.2e-04 & 1.3e-04 \\ & & -0.5005 & 0.7004 & -0.4997 & 0.6990 & 6.2e-04 & 5.8e-04 & 1.3e-04 & 1.2e-04\\ \cline{2-10}
     & 2000 & 0.9001 & 0.5012 & 0.8998 & 0.4999 & 3.4e-04 & 3.7e-04 & 6.2e-05 & 5.8e-05 \\ & & -0.4997 & 0.6998 & -0.4997 & 0.6995 & 3.0e-04 & 2.8e-04 & 6.1e-05 & 6.0e-05\\ 
    \hline
\end{tabular}
\end{table*}

\begin{table*}[t]
\centering
\caption{Mean squared error of the different estimators used in the threshold-free scheme for three \VAR{2}{1} and different numbers of samples. For all the configurations, the input noise has a standard Gaussian distribution and $10^3$ simulations were made to compute the average and MSE matrices. The last column corresponds to the expected variance for $\hat A_1$ if the estimations of the matrix $\tilde A_1$ and of the variance ratios would have been independent.}
\label{tab:VAR2-1_perf2}
\begin{tabular}{|c|r|r r|r r|r r|r r|}
    \hline\rule{0em}{1.25em}
    $A_1$ & $T$ & \multicolumn{2}{c|}{MSE $\hat A_1$} &\multicolumn{2}{c|}{MSE $\hat{ \tilde A}_1$} &\multicolumn{2}{c|}{MSE $\hat r_{i,j}$} &\multicolumn{2}{c|}{var if independent}\\
    \hline
    \multirow{8}{*}{$\begin{pmatrix} 0.25 & 1.00 \\ 0.00 & -0.20 \end{pmatrix}$}& 250 & 7.9e-03 & 1.7e-02 & 7.9e-03 & 4.7e-03 & - & 1.7e-02 & 7.9e-03 & 1.8e-02 \\ & & 5.8e-03 & 1.1e-02 & 1.1e-02 & 1.1e-02 & 4.5e-03 & - & 5.7e-03 & 1.1e-02\\ \cline{2-10}
     & 500 & 4.0e-03 & 8.2e-03 & 4.0e-03 & 2.4e-03 & - & 8.3e-03 & 4.0e-03 & 8.9e-03 \\ & & 2.8e-03 & 5.2e-03 & 5.5e-03 & 5.2e-03 & 2.1e-03 & - & 2.8e-03 & 5.2e-03\\ \cline{2-10}
     & 1000 & 1.9e-03 & 4.3e-03 & 1.9e-03 & 1.2e-03 & - & 4.3e-03 & 1.9e-03 & 4.5e-03 \\ & & 1.3e-03 & 2.4e-03 & 2.6e-03 & 2.4e-03 & 1.1e-03 & - & 1.3e-03 & 2.4e-03\\ \cline{2-10}
     & 2000 & 9.4e-04 & 2.1e-03 & 9.4e-04 & 5.6e-04 & - & 2.1e-03 & 9.4e-04 & 2.2e-03 \\ & & 6.8e-04 & 1.1e-03 & 1.4e-03 & 1.1e-03 & 5.4e-04 & - & 6.8e-04 & 1.1e-03\\ 
    \hline
    \multirow{8}{*}{$\begin{pmatrix} 0.70 & 0.00 \\ 1.00 & -0.50 \end{pmatrix}$}& 250 & 8.7e-03 & 9.6e-03 & 8.7e-03 & 1.3e-02 & - & 6.9e-03 & 8.7e-03 & 9.6e-03 \\ & & 1.7e-02 & 1.0e-02 & 1.0e-02 & 1.0e-02 & 1.3e-02 & - & 2.3e-02 & 1.0e-02\\ \cline{2-10}
     & 500 & 4.3e-03 & 4.7e-03 & 4.3e-03 & 6.2e-03 & - & 3.3e-03 & 4.3e-03 & 4.7e-03 \\ & & 7.6e-03 & 5.2e-03 & 4.9e-03 & 5.2e-03 & 5.9e-03 & - & 1.1e-02 & 5.2e-03\\ \cline{2-10}
     & 1000 & 2.1e-03 & 2.2e-03 & 2.1e-03 & 3.0e-03 & - & 1.8e-03 & 2.1e-03 & 2.2e-03 \\ & & 3.7e-03 & 2.5e-03 & 2.5e-03 & 2.5e-03 & 3.1e-03 & - & 5.7e-03 & 2.5e-03\\ \cline{2-10}
     & 2000 & 1.1e-03 & 1.1e-03 & 1.1e-03 & 1.4e-03 & - & 9.4e-04 & 1.1e-03 & 1.1e-03 \\ & & 1.9e-03 & 1.2e-03 & 1.2e-03 & 1.2e-03 & 1.6e-03 & - & 2.8e-03 & 1.2e-03\\ 
    \hline
    \multirow{8}{*}{$\begin{pmatrix} 0.90 & 0.50 \\ -0.50 & 0.70 \end{pmatrix}$}& 250 & 2.6e-03 & 3.1e-03 & 2.6e-03 & 1.6e-03 & - & 5.7e-03 & 2.6e-03 & 3.1e-03 \\ & & 2.6e-03 & 2.2e-03 & 1.2e-03 & 2.2e-03 & 5.2e-03 & - & 2.5e-03 & 2.2e-03\\ \cline{2-10}
     & 500 & 1.3e-03 & 1.5e-03 & 1.3e-03 & 8.8e-04 & - & 2.7e-03 & 1.3e-03 & 1.6e-03 \\ & & 1.2e-03 & 1.1e-03 & 5.7e-04 & 1.1e-03 & 2.5e-03 & - & 1.2e-03 & 1.1e-03\\ \cline{2-10}
     & 1000 & 6.0e-04 & 7.7e-04 & 6.0e-04 & 4.5e-04 & - & 1.4e-03 & 6.0e-04 & 8.0e-04 \\ & & 6.2e-04 & 5.8e-04 & 2.7e-04 & 5.8e-04 & 1.3e-03 & - & 6.0e-04 & 5.8e-04\\ \cline{2-10}
     & 2000 & 3.4e-04 & 3.7e-04 & 3.4e-04 & 2.2e-04 & - & 6.8e-04 & 3.4e-04 & 3.9e-04 \\ & & 3.0e-04 & 2.8e-04 & 1.4e-04 & 2.8e-04 & 6.3e-04 & - & 3.0e-04 & 2.8e-04\\ 
    \hline
\end{tabular}
\end{table*}

\subsection{Higher dimension: Use of redundancy in $\vQ(t)$}

This part highlights how the redundancy of the measure $\vQ(t)$ in the threshold-free scheme can be exploited either to increase precision or to reduce the amount of data. Three variant schemes have been implemented: a simple scheme that does not benefit from this redundancy, an optimized scheme that uses it to increase precision and an efficient scheme that requires fewer data. The difference between the three solutions comes from the computation of the ratios $\hat r_{i,j}$:
\begin{itemize}
    \item the simple scheme follows the algorithm presented in part \ref{sec:Scheme2}, each ratio $\hat r_{i,j}^{(s)}$ is obtained using only the corresponding measure of preponderance $Q_{i,j}(t)$;
    \item the optimized approach takes advantage of the redundancy to optimize the ratios: it first estimates the $r_{i,j}^{(s)}$ using the preceding method, then it computes the $\left\{\hat \sigma_i\right\}_i$ which minimize $l(\vsigma)=\sum (\sigma_i - \hat r_{i,j}^{(s)} \sigma_j)^2$ constrained to $||\vsigma|| = 1$ --where the sum is over all the couples $(i,j) \in \intEnt{1,d}^2$ --the ratios are finally defined as $\hat r_{i,j}^{(o)}=\hat \sigma_i/\hat \sigma_j$;
    \item the efficient approach removes redundancy to cut the need of data: the ratio $\hat r_{i,j}^{(e)}$ is deduced from the $d-1$ ratios $\left\{\hat r_{i,1}^{(s)}\right\}_{i\ge 2}$ with the formula: $\hat r_{i,j}^{(e)} = \hat r_{i,1}^{(s)} / \hat r_{j,1}^{(s)}$.
\end{itemize}

The optimized scheme uses the redundancy to improve the estimation of the standard deviations. Note that the constraint $||\vsigma|| = 1$ prevents to get the trivial solution $\vsigma = \vzeros$ but the constant is arbitrary. This optimization problem is equivalent to finding the eigenvector associated to the smallest eigenvalue of a specific matrix. It can be proved (cf. section \ref{proof:ratios_optimization}) that all the components of $\hat \vsigma$ have the same sign and are not null: consequently, the optimized ratios are well defined and positive. Another convenient solution could have been to minimize $l'(\vtheta) = \sum (\theta_i - \theta_j - \log r_{i,j})^2$ and then to define $\hat \sigma_i = \exp(\theta_i)$. Doing this way the estimated standard deviation would have also been clearly strictly positive.

The efficient scheme uses the dependence between the ratio to reduce the number of unknown parameters. As all ratios can be deduced from the ratios with the first series, recording only the comparisons with this particular time series is enough. This result is very useful in practise, as it implies that not all the sensors need to be connected. We can imagine a sensor network in which the sensors can  communicate  only with their neighbours for proximity reason. In such a case, a new optimization method would have to be designed to deduce the missing ratios and take advantage of possible redundancy. 

To compare these three schemes, we generated $10^3$ random \VAR{d}{1} of several dimensions $d \in \intEnt{2,8}$ by an acceptance-rejection method--a model was accepted if its largest eigenvalue modulus was between $0.5$ and $0.85$ (note that the distribution of the eigenvalues is not uniform and may change with $d$). Table \ref{tab:VARd-1_perf} reports the MSE for the three methods compared to the MLSE on the continuous data. First of all, we note that as expected, in the bivariate case, the three methods are perfectly equivalent. For higher dimensions, the optimized scheme is more accurate than the simple scheme which is more accurate than the efficient scheme. This was predictable as the efficient scheme uses less data. Interestingly enough, we can observe that reducing at maximum the amount of data does not induce a significant increase of the error: the error increases by less than $5\%$ ($2\%$ in average) between the non-diagonal parameters of the optimized scheme and the non-diagonal and non-first-time-series-related parameters of the data efficient scheme. This result must be nuanced, in this ideal case, all the sensors were connected to a same sensor so at worst only two ratios were used to deduce any ratio, if for some reason, the sensors can communicate only with their neighbours, this number can be greater: for example to compute $\hat \sigma_1/\hat\sigma_4$ a three-term chained rule must be used: $\sigma_1/\sigma_4 = \sigma_1/\sigma_2 \times \sigma_2/\sigma_3 \times \sigma_3/\sigma_4$ and as the each ratio estimation is not perfectly accurate, the longer the chain, the less accurate.

\begin{table*}
\centering
\caption{Mean squared error of the three threshold-free estimators introduced and the MLSE on the continuous data for \VAR{}{1} models of different dimensions. For all the configurations, the input noise has a standard Gaussian distribution, $2000$ samples were used and $10^3$ random models were simulated to compute the average. For all the methods, the MSE on the non-diagonal coefficients is noted in brackets--for the efficient method we distinguished the coefficients of the first raw or column (first value) from the others (second value).}
\label{tab:VARd-1_perf}
\begin{tabular}{|c|c|c|c|c|}
    \hline
    $d$ & Simple Scheme & Optimized Scheme & Efficient Scheme & Cont. MLSE\\
    \hline
    2 & 7.26e-03 (1.61e-02) & 7.26e-03 (1.61e-02) & 7.26e-03 (1.61e-02 / -) & 4.08e-04 (8.16e-04) \\
	3 & 7.60e-03 (1.56e-02) & 7.57e-03 (1.55e-02) & 7.64e-03 (1.55e-02 / 1.62e-02) & 3.99e-04 (8.03e-04) \\
	4 & 4.53e-03 (9.11e-03) & 4.52e-03 (9.07e-03) & 4.55e-03 (9.05e-03 / 9.26e-03) & 3.91e-04 (7.85e-04) \\
	5 & 4.14e-03 (8.17e-03) & 4.12e-03 (8.13e-03) & 4.16e-03 (8.25e-03 / 8.22e-03) & 3.87e-04 (7.73e-04) \\
	6 & 3.93e-03 (7.75e-03) & 3.91e-03 (7.72e-03) & 3.94e-03 (7.79e-03 / 7.80e-03) & 3.89e-04 (7.80e-04) \\
	7 & 3.83e-03 (7.66e-03) & 3.81e-03 (7.63e-03) & 3.85e-03 (7.65e-03 / 7.73e-03) & 3.88e-04 (7.74e-04) \\
	8 & 3.78e-03 (7.56e-03) & 3.76e-03 (7.53e-03) & 3.79e-03 (7.59e-03 / 7.60e-03) & 3.97e-04 (7.92e-04) \\
    \hline
\end{tabular}
\end{table*}

\section{Concluding remarks}

The main contribution of this paper is the introduction of two algorithms for identifying the parameters of a VAR model from binary quantised measurements. The first method uses asymmetric quantisation while the second method uses symmetric quantisation. The second method also requires quantised measurements of the pairwise differences of the model components.

From an applications perspective, these two methods complement each other. The first method is more data efficient --- $d$ bits are recorded at each time step compared to $2d-1$ in the most efficient implementation of the second method --- and does not require any communication between pairs of sensors. Good performance requires the quantisation threshold to be chosen appropriately, as a function of the variance. For some applications, the variance is known \textit{a priori} and this is not a limitation. Although beyond the scope of this paper, we envisage an adaptive method for optimising the threshold can be developed for applications where the variance is unknown and may even change over time.

The second method removes the need to choose a threshold but requires more measurements. These extra measurements require the instantaneous raw sensor outputs to be compared with each other to form a ranking. The benefit though over the first method is there is no need to estimate the component variances beforehand.

Interestingly, the very coarse 1-bit quantisation does not affect the asymptotic performance of the parameter estimates: like in the standard case, the variances of the parameter estimates decrease at the rate $1/T$ where $T$ is the number of vector observations. Moving to 1-bit quantisation has the potential to reduce networking and energy costs considerably.

\section{Proofs}

\subsection{Predominance probability}\label{proof:predominace}

Let $Z_1$ and $Z_2$ be two zero mean Gaussian random variables with respective variance $\sigma_1^2$ and $\sigma_2^2$. We denote the correlation between $Z_1$ and $Z_2$ as $\rho$, the ratio $\sigma_1/\sigma_2$ as  $r$ and the probability $\prob\left(| Z_1 | \ge | Z_2 | \right)$ as  $p$ . We will prove:
\begin{align}
    p &= \frac{1}{\pi} \left[ \arctan\left(\frac{r-\rho}{\sqrt{1-\rho^2}}\right) + \arctan\left(\frac{r+\rho}{\sqrt{1-\rho^2}}\right) \right] \\
    r &= \sqrt{\frac{1-\rho^2}{\tan(\pi p)^2} + 1} - \frac{\sqrt{1-\rho^2}}{\tan(\pi p)}
\end{align}

We start by decomposing $p$ as:
\begin{align*}
    p &= 2\prob\left( Z_1 \ge |Z_2| \right) \\
    &= 2\prob\left( Z_1 \ge Z_2 \mid Z_2 > 0\right) + 2\prob\left( Z_1 \ge -Z_2 \mid Z_2 < 0\right) 
\end{align*}
We then perform a change of variables to normalize and decorrelate $Z_1$ and $Z_2$ (replace $\rho$ by $-\rho$ for $Z_1$ and $-Z_2$):
\begin{equation*}
    \begin{pmatrix}U_1 \\ U_2\end{pmatrix} = \begin{pmatrix} \frac{1}{\sigma_1} & 0 \\ \frac{-\rho}{\sigma_1\sqrt{1-\rho^2}} & \frac{1}{\sigma_2\sqrt{1-\rho^2}}\end{pmatrix} \begin{pmatrix}Z_1 \\ Z_2\end{pmatrix}
\end{equation*}
We obtain:
\begin{align*}
    \prob\left( Z_1 \ge Z_2 \mid Z_2 > 0\right) &= \int_{0}^{+\infty}\int_{-\frac{\rho u_1}{\sqrt{1-\rho^2}}}^{\frac{(r-\rho)u_1}{\sqrt{1-\rho^2}}} \phi_2(u_1, u_2) du_2 du_1 \\
    \prob\left( Z_1 \ge -Z_2 \mid Z_2 < 0\right) &= \int_{0}^{+\infty}\int_{\frac{\rho u_1}{\sqrt{1-\rho^2}}}^{\frac{(r+\rho)u_1}{\sqrt{1-\rho^2}}} \phi_2(u_1, u_2) du_2 du_1
\end{align*}
where $\phi_2$ is the joint PDF of a standard bivariate Gaussian variable. We conclude by noticing that, in the previous formulas, the integrals are over a cone which leads to
\begin{multline*}
    \pi p = \arctan\left(\frac{r-\rho}{\sqrt{1-\rho^2}}\right) - \arctan\left(-\frac{\rho}{\sqrt{1-\rho^2}}\right) +\\
    \arctan\left(\frac{r+\rho}{\sqrt{1-\rho^2}}\right) - \arctan\left(\frac{\rho}{\sqrt{1-\rho^2}}\right)
\end{multline*}
and we obtain lemma \ref{lem:ratios}.

To invert this formula, we used the well known formula on the sum of arctangents and we obtain:
\begin{equation*}
    \pi (p+n) = \arctan\left(\frac{2r\sqrt{1-\rho^2}}{1-r^2}\right)
\end{equation*}
where $n \in \Z$.

By taking the tangent and solving the second order equation (as $r > 0$), we find the announced formula.

\subsection{Optimization of the ratios}\label{proof:ratios_optimization}

Let $R = \left(r_{i,j}\right)_{1\le i,j\le d}$ be a $d$-shaped square matrix with strictly positive coefficients and $1$s on its diagonal. More over, let $l$ be the penalization function:
\begin{equation}
    l:\left\{\begin{array}{ccc}
         \R^d &\longrightarrow & \R  \\
         \vtheta& \longmapsto & \sum\limits_{1 \le i,j\le d} (\theta_i - r_{i,j} \theta_j)^2
    \end{array}\right.
\end{equation}
We will first solve the problem:
\begin{equation}
    \hat \vtheta = \argmin{\vtheta \in S^d} l(\theta)
\end{equation}
where $S^d$ denote the unit d-sphere.
Secondly, we will prove that all the components of the solutions have the same sign and are not null.

As $l$ is a quadratic form, it can be reformulated using a $d$-shaped symmetric matrix $A$ such that for all $\vtheta$, $l(\vtheta) = \vtheta^\intercal A \vtheta$.
\begin{equation}
    A = \left(a_{i,j}\right)_{1 \le i,j \le d} = \left\{ \begin{array}{clc}
         (d-2) + \sum\limits_{k=1}^d r_{k,i}^2 & \text{if} & i=j \\
         -(r_{i,j} + r_{j,i})& \text{if}& i\ne j
    \end{array}\right.
\end{equation}
As $A$ is a real-valued symmetric matrix, it is diagonalizable and consequently the solutions of the problem are the normalized eigenvectors associated to the smallest eigenvalue of $A$.

Let $\vxi$ be a such eigenvector, let us prove that all its components have the same sign. Let $\tilde \vxi = \begin{pmatrix} | \xi_{1} | & \cdots & | \xi_{d} |\end{pmatrix}^\intercal$ be the vector formed by the absolute values of the $\vxi$'s components,
\begin{align*}
    l(\vxi) &= \sum_{i=1}^d \left[(d-2)+\sum_{k=1}^d r_{k,i}^2\right]\xi_i^2 - \sum_{i\ne j}(r_{i,j}+r_{j,i})\xi_i \xi_j\\
    &\ge \sum_{i=1}^d \left[(d-2)+\sum_{k=1}^d r_{k,i}^2\right]\xi_i^2 - \sum_{i\ne j}(r_{i,j}+r_{j,i}) | \xi_i \xi_j |\\
    &= l(\tilde \vxi)
\end{align*}
But since $\tilde \vxi \in S^d$, $l(\tilde \vxi) \ge l(\vxi)$ so $l(\tilde \vxi) = l(\vxi)$. As the $r_{i,j}$ are strictly positive, the case equality occurred iff $| \xi_i \xi_j | = \xi_i \xi_j$ \ie all the components of $\vxi$ have the same sign.

Finally, we prove by contradiction that all the components of $\vxi$ are not null. We assume w.l.o.g. that all the components of $\vxi$ are positive and that $\xi_1 =0$ and $\xi_2 > 0$. Let $\vxi' = \begin{pmatrix}\varepsilon & \sqrt{\xi_2^2 - \varepsilon^2} & \xi_3 & \dots & \xi_d\end{pmatrix}^\intercal \in S^d$ with $0 < \varepsilon < \xi_2$. We have $l(\vxi') \ge l(\vxi)$ and:
\begin{equation}
    l(\vxi') = l(\vxi) - 2\varepsilon\sum_{j=2}^d (r_{1,j} + r_{j,1})\xi_j + o(\varepsilon)
\end{equation}
So, it exists $\varepsilon > 0$ such that $l(\vxi') < l(\vxi)$ which is impossible.

\subsection{Graph representation}\label{proof:graph_representation}

Let $\rG = (V,E)$ be the graph representing the sensor network. We will give an expression of the variance of the estimation of $a_{i,j}^{(s)}$, the entry at the $i$-th row and the $j$-th column in $A_s$. To do so, we will assume:
\begin{itemize}
    \item the estimation of the ratios $r_{i,j}$ is independent of the estimation of the unscaled parameters $\tilde a_{i,j}^{(s)}$, this assumption has already been discussed;
    \item the estimation of the ratios are mutually independent, this is wrong but it allows to explain the behaviour;
    \item the variance $\sigma_{\tilde a}^2$ of the estimation of the unscaled parameters is the same for all the parameters and so is the variance $\sigma_{r}^2$ of the ratios compute directly from the measurements (\ie without using the chained rule) --this is generally wrong but as they are close we chose to identify them to simplify notations.
\end{itemize}

We consider a couple of sensors $(i,j)$ and note as $d(i,j)$ the distance between $i$ and $j$ in the graph, we will consider the path used to estimate $r_{i,j}$ \ie the sequence of vertices $\rI_{i,j} = \{k_0=i, k_1, \dots, k_{d(i,j)} = j\}$ such that $\forall l \in \intEnt{1,d(i,j)}$, $(k_{l-1},k_l) \in E$. As $a_{i,j}^{(s)} = r_{i,j} \tilde a_{i,j}^{(s)}$, using independence we obtain:
\begin{equation}
    \var(\hat a_{i,j}^{(s)}) = \var(\hat r_{i,j}) \left[\sigma_{\tilde a}^2 + \left(\tilde a_{i,j}^{(s)}\right)^2 \right] + r_{i,j}^2 \sigma_{\tilde a}^2
\end{equation}

More over as the  $\hat r_{i,j} = \prod_{l=1}^{d(i,j)}\hat r_{k_{l-1},r_l}$, using independence, we have:
\begin{align*}
    \var(\hat r_{i,j}) &= \E\left( \prod_{l=1}^{d(i,j)}\hat r_{k_{l-1},k_l}^2\right) - \E\left( \prod_{l=1}^{d(i,j)}\hat r_{k_{l-1},k_l}\right)^2 \\
    &= \prod_{l=1}^{d(i,j)}(\sigma_r^2 + r_{k_{l-1},k_l}^2) - \prod_{l=1}^{d(i,j)} r_{k_{l-1},k_l}^2 \\
    &= \sigma_r^2 \sum_{l=1}^{d(i,j)} \left(\frac{1}{r_{k_{l-1},k_l}^2} \prod_{m=1}^{d(i,j)} r_{k_{m-1},k_m}^2\right) + o(\sigma_r^2)\\
    &= \sigma_r^2 r_{i,j}^2 \sum_{l=1}^{d(i,j)} r_{k_l,k_{l-1}}^2 + o(\sigma_r^2)
\end{align*}

By injecting this result in the previous formula, we decomposed the variance into a term independent of the graph $r_{i,j}^2 \sigma_{\tilde a}^2$ and a term related to the path taken to estimate $r_{i,j}$. Note that as expected, if $d(i,j) = 0$ (\ie $i=j$), we found that $\var(\hat r_{i,j}) = 0$ and if $d(i,j)=1$ (\ie the $i$-th and the $j$-th sensor can communicate), $\var(\hat r_{i,j}) = \sigma_r^2$.

\bibliographystyle{unsrt}

\end{document}